\documentclass[prl,aps,twocolumn,floats,showpacs]{revtex4}
\usepackage{graphicx}
\usepackage{psfrag}

\begin{document}

\title{Violation of Bell's inequality: criterion for quantum communication complexity advantage}

\author{{\v C}aslav Brukner$^1$, Marek {\. Z}ukowski$^{1,2}$, Jian-Wei Pan$^1$, Anton Zeilinger$^1$}
\affiliation{$^1$Institut f\"ur Experimentalphysik, Universit\"at Wien, Boltzmanngasse 5, A--1090 \\
$^2$Institute Fizyki Teoretycznej i Astrofizyki Uniwersytet
Gda\'nski, PL-80-952 Gda\'nsk, Poland}

\date{\today}

\begin{abstract}

We prove that for every Bell's inequality and for a
broad class of protocols, there always exists a multi-party
communication complexity problem, for which the protocol assisted
by states which violate the inequality is more efficient
than any classical protocol. Moreover, for that advantage Bell's
inequality violation is a necessary and sufficient criterion.
Thus, violation of Bell's inequalities has a significance beyond
that of a non-optimal-witness of non-separability.

\end{abstract}

\pacs{3.65 Bz, 3.67 -a, 42.50 Ar}

\maketitle

Entanglement is the essential feature, which distinguishes the
quantum  from the classical \cite{schroedinger}. On one hand,
entangled states violate Bell inequalities, and thus rule out
local realistic explanation of quantum mechanics \cite{bell}. On
the other hand, they enable certain communication and
computation tasks to have an efficiency not achievable by the laws of
classical physics \cite{nielsen}.

Intuition suggests that these two aspects, the fundamental one,
and the applicational one, could be intimately linked.
Specifically, one could expect, that only the quantum communication protocol which
makes use of an entangled state which violates some Bell's
inequality can have efficiency larger than any classical protocol. 
Otherwise one might expect that the efficiency of the protocol could be explainable by
a local realistic model, and thus achievable in classical
physics. This intuitive reasoning is supported by the result of
Ref. \cite{scarani} where it was shown that violation of Bell's inequality
is a condition for the security of quantum key
distribution protocols. Here we give another result which supports the intuitive reasoning: 
the violation of Bell's inequalities is a necessary and sufficient criterion for
the quantum communication complexity protocol to be
more efficient than any classical one. 

We shall discuss the following version of the communication
complexity problems (such problems were introduced in Ref.
\cite{yao}). Some input data are distributed over $n$ separated
parties. Every party knows the local data, but not the data of
the others. The party $i$ obtains an input string $z_i$. The goal
is for each of them to determine the value of some function
$f(z_1,...,z_n)$, while exchanging a {\it restricted} amount of
information. This restriction, in general, enables the parties to
compute the function only with an error. Then the goal for all parties
is to compute the function correctly with as high a probability as
possible. An execution is considered successful, if the values
determined by {\it all} parties are correct. Before they start
the protocol, the parties are allowed to share (classically) correlated random
strings, or any other data, which might improve the success of
the protocols. They are allowed to process their data locally in
whatever way.

The general question is whether and to what extent entanglement
can be of advantage for solving such problems. It
was shown that entanglement can improve the probability of
success in communication complexity protocols beyond the limits
which are classically possible
\cite{buhrman,buhrman1,brassard,cleve,grover1}. Specifically, Buhrman, Cleve
and van Dam \cite{buhrman} found a two-party communication
complexity problem, that can be solved with a higher probability
of success than in any classical protocol, if the parties share
entanglement. The quantum protocol was based on violation of the
Clauser-Horne-Shimony-Holt (CHSH) inequality \cite{chsh} by
two-qubit maximally entangled state. Similarly, the quantum
protocols of multi-party problems of Ref. \cite{buhrman,buhrman1}
were based on an application of the GHZ-type \cite{ghz} argument
against local realism for multi-qubit maximally entangled states.
In Ref. \cite{galvao} an equivalence between the CHSH and GHZ
tests for three particles and the two- and three-party quantum
protocols of Ref. \cite{buhrman}, respectively, was shown. All
these results indicate that there is a link between the quantum
communication complexity protocols and the violation of Bell's
inequality.

However, the problems: (a) which quantum states are needed to
achieve an improvement of the probability of success over any
classical strategy and (b) what are the classes of functions for
which this improvement is possible, are still open. We address
the question (a) and (b) for a class of $n$-party communication
complexity problems. We prove, that for any Bell's inequality for
$n$ qubits  one can formulate at least one communication complexity problem
with the following property. The success of the quantum protocol (i.e. which uses entangled states)
is higher than in any classical protocol {\it if and only if} the $n$ qubits
violate the Bell inequality. As exemplary applications we use the
complete set of $2^{2^n}$ of $n$-qubit Bell's inequalities for
correlation functions \cite{ww,zb}. For this example we find a
family of functions for which the improvement of the probability of
success over classical strategies is possible. We also present
extensions of these results.

Let us define the general $n$-party communication complexity
problems to be considered:
\begin{itemize}
\item The $i$-th party receives a two-bit input string
$(x_i, y_i)$. (For convenience the values of the bits are encoded
as follows $x_i=0$ or $1$, and $y_i=-1$ or $1$.)
\item The distributed values for $y_i$ are
chosen randomly and for the $x_i$'s in accordance with a certain
probability distribution $Q(x_1,...,x_n)$. Thus the inputs $x_1,...,x_n$ can be
(classically) correlated.
\item After receiving the input strings each party is allowed to
broadcast only {\it one bit} of information (denoted as $e_i$).
It may reveal e.g. a part of the received string, or some locally
produced result of computation or measurement.
\item Finally each party attempts to give a value for the function
$f(x_1,...,x_n,y_1,...,y_n)$, with $f={\pm1}$. The execution of the
protocol is successful when {\em all} parties arrive at the
correct value of $f$. Their joint task is to maximize the probability
of success.
\end{itemize}

We shall consider a specific sub-class of the above problems, for
which there exists a real-valued function $g(x_1,...,x_n)$, such
that
\begin{equation}
Q(x_1,...,x_n)=\frac{|g(x_1,...,x_n)|}{\sum_{x_1,...,x_n=0}^{1}|g(x_1,...,x_n)|},
\label{prior}
\end{equation}
and
\begin{equation}
f= y_1 \!\cdot\! y_2 \cdot ... \cdot y_n \cdot S[g(x_1,x_2,...,x_n)],
\label{function}
\end{equation}
where $ S[g]=g/|g| = \pm 1$ is the sign function of $g$.

We shall prove that for any Bell's inequality for qubits there
exists at least one problem from the above class, such that the
probability of success in the quantum protocol (i.e. which uses an
entangled state) is higher than in any classical one. This is so
if, and only if, the entangled state used violates the Bell inequality
for correlation functions
\begin{equation}
\sum_{x_1,...,x_n=0}^{1} g(x_1,...,x_n) E({x_1},...,{x_n}) \leq
B(n),
\label{vatra}
\end{equation}
(this general form includes also inequalities not known yet).
In Ineq. (\ref{vatra})
$E({x_1},...,{x_n})$ is a shorthand notation for the Bell-type
correlation function $E(O^1_{x_1},...,O^n_{x_n})$, 
for measurements on $n$ particles, which involve, at each
local measurement station $i$, two alternative dichotomic
observables $O^i_{0}$ and $O^i_{1}$, each of spectrum $\pm 1$.

We now present a broad class of classical protocols which will be
considered here:
\begin{enumerate}
\item
Each party $i$ calculates (e.g., with help of a computer) locally
any function $a_i(x_i, \lambda_i)$, where $\lambda_i$ is any
other parameter, or a set of parameters, on which the function
$a_i$ may additionally depend. For example, $\lambda_i$ can be  a random string
of variables shared among the parties before they start the
protocol. Each party $i$ broadcasts $e_i\!=\!a_i\cdot y_i$.
\item After the broadcast all parties put as the value of $f$ the number 
$y_1 \cdot ... \cdot y_n \cdot  a_1 \cdot ... \cdot a_n $,
which is equal to the actual value of function $f$ for a certain fraction of
cases (see below).
\end{enumerate}

Let us calculate the probability of success achievable for the
considered class of classical protocols. It is equal to the
probability, $P$, that the product $a_1 \cdot ... \cdot a_n$ of
the locally computed functions is equal to $S[g]$:
\begin{equation}
P \!= \hspace{-0.3cm} \sum_{x_1,...,x_n=0}^{1}
\hspace{-0.1cm}  Q(x_1,...,x_n) \cdot P_{{x_1}...{x_n}} \left(a_1
\cdot ... \cdot a_n \!= \!S[g] \right), \label{sah}
\end{equation}
where $P_{{x_1}...{x_n}} \left(a_1 \cdot ...\cdot a_n = S[g]
\right) $ is the probability that $a_1 \cdot ... \cdot a_n =
S[g(x_1,...,x_n)]$ if parties receive inputs ${x_1,...,x_n}$.

Next, we introduce a quantum competitor of the class of classical
protocols considered above. The parties share $n$ entangled
qubits. Each of them can perform measurements on the local qubit.
The quantum protocol reads (Fig. 1):
\begin{enumerate}
\item If party $i$ receives $x_i=0$, she will measure her qubit
with the local  apparatus, which is set to measure a dichotomic
observable $O^i_0$. Otherwise, i.e. for $x_i=1$, she  measures a
different  observable $O^i_1$. We ascribe to the outcomes of the
measurements the two values $\pm 1$. The actual value obtained by
party $i$ will be denoted as $a_i$. It will serve the same role
as the result of local computation in the classical protocol.
Each party $i$ broadcasts  $e_i=a_i \cdot y_i$.
\item After the broadcast all parties put as the value of $f$ the number 
$y_1 \cdot ... \cdot y_n \cdot  a_1 \cdot ... \cdot a_n $,
which is equal to the actual value of function $f$ for a certain fraction of
cases.
\end{enumerate}

The probability of success in the quantum protocol is the
probability, $P$, that the product $\prod_{i=1}^n a_i$ of
the local measurement results is equal to $S[g]$. Thus, it
can also be expressed by Eq. (\ref{sah}) where now
$P_{{x_1}...{x_n}} \left(a_1 \cdot ...\cdot a_n = S[g] \right) $
is the probability that $\prod_i a_i =
S[g(x_1,...,x_n)]$ if parties  measure their qubits with the local
apparatus set at $O^1_{x_1},..., O^n_{x_n}$.

\begin{figure}
\centering
\includegraphics[angle=0,width=8.9cm]{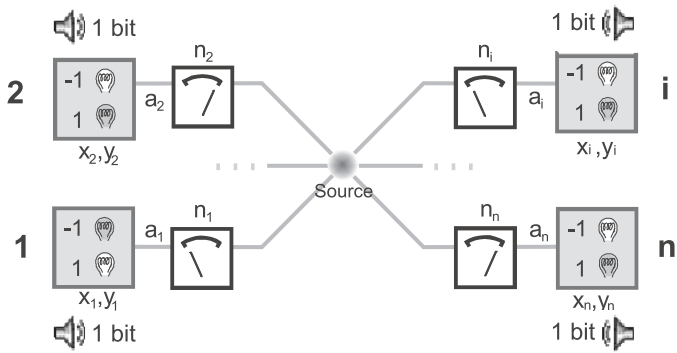}
\caption{Multi-party quantum communication complexity protocol
which is based on the Bell experiment with $n$ qubits. Every
party $i$ receives an input string $(x_i,y_i)$ where both $x_i$
and $y_i$ are bit values. Depending on the values of $x_i$ party $i$
chooses to measure between two different two-values
observables $O^i_0$ or $O^i_1$. The actual measurement result
obtained by party $i$ is denoted by $a_i$. Each party broadcasts
the product $y_i\cdot a_i$.}
\label{multibellexp} \vspace{-0.5cm}
\end{figure}

It is essential to realize that the classical protocols introduced
above are equivalent to a local realistic model of the quantum
protocol because $\lambda$'s can be considered as local hidden
variables, which can be shared between parties. We will now show that the
combination of probabilities on the right-hand side of Eq.
(\ref{sah}), in the case
of a classical protocol, is bounded by the limits
imposed by local realistic models. That is, the combination of
probabilities in Eq. (\ref{sah}) satisfies a Bell-type inequality.

Note that the correlation function is given by $E\! = P(\prod_i
a_i=1)- P(\prod_i a_i =-1) $ and therefore
\begin{equation} E({x_1},..,{x_n})\! =
S[g] \big(2
P_{{x_1}...{x_n}}(a_1 \! \cdot \! ... \! \cdot a_n \!=\!S[g]) -1 \big).
\label{mrav}
\end{equation}
Using this one easily shows that the right hand side of Eq.
(\ref{sah}) is proportional to the left hand side of Ineq.
(\ref{vatra}). One obtains
\begin{eqnarray}
\sum_{x_1,...,x_n} Q(x_1,...,x_n) \cdot
P_{{x_1}...{x_n}}\left(a_1 \cdot ... \cdot a_n\!=\! S[g]\right) \leq  \nonumber \\
\frac{1}{2}\left(1 + \frac{B(n)}{\sum |g(x_1,...,x_n)|}\right).
\label{mrsko}
\end{eqnarray}
If Ineq. (\ref{vatra}) is violated, so is Ineq. (\ref{mrsko}), and
vice versa. That is, the entanglement assisted protocol can
result in a higher probability of a success than any classical
one of the considered class, {\it if and only if} the respective Bell's
inequality is violated \cite{remark1}.

Let us now present some examples. The set of $2^{2^n}$ Bell's
inequalities of the form (\ref{vatra}) was obtained in Ref.
\cite{ww,zb}. There the class of functions $g$ is given by
\begin{equation}
g(x_1,...,x_n)= \hspace{-2mm}
\sum_{s_1,...,s_n=-1}^{1} S(s_1,...,s_n) \cdot
s^{x_1}_1 \cdot ... \cdot s^{x_n}_n,
\label{class}
\end{equation}
where $S(s_1,...,s_n)\!=\! \pm 1 $ is a sign function and the
bound is $B(n) \!=\! 2^n$. There are $2^{2^n}$ different sign
functions and correspondingly $2^{2^n}$ different functions $g$.
The functions, which can lead to a greater success probability in the 
case of quantum protocols, are those which are associated with non-trivial Bell inequalities
of the form (\ref{class}), that is such ones which are violated by 
quantum predictions. Note that
factorable functions,
 $S(s_1,...,s_n)= S_1(s_1)\cdot...\cdot S_n(s_n)$ are therefore excluded from this family. In
the following we give two explicit functions $g$ of the class
(\ref{class}). We only give the final results, as these follow
from the general proof given above.

Consider $S_{odd}= \sqrt{2}
\cos\left[(s_1\!+\!...\!+\!s_n)\frac{\pi}{4}\right]$ for $n$ odd
and $S_{even}=
\cos\left[(s_1\!+\!...\!+\!s_n)\frac{\pi}{4}\right]$ for $n$ even.
This implies for $n$ odd
\begin{equation}
g_{odd}=\sqrt{2^{n+1}} \cos\left[\frac{\pi}{2}(x_1 \! +
\!x_2\!+\!...\!+\! x_n)\right].
\end{equation}
whereas for $n$ even one has
\begin{equation}
g_{even}=\sqrt{2^{n}} \cos\left[\frac{\pi}{2}(x_1 \! +
\!x_2\!+\!...\!+\! x_n)\right].
\end{equation}
The probability distribution  $Q(x_1,...,x_n)$ is such that with
equal probability only the input strings $x_i$ which satisfy the
condition that $x_1+...+x_n$ is even are distributed. This
specific type of  problem was first considered by Buhrman {\it et
al} \cite{buhrman,buhrman1}. The quantum protocol rests on the
violation of an inequality, which is equivalent to the Mermin
inequality \cite{mermin,belinskii}. The maximal probability of
success in the classical protocol is $P^{max}_C\!=\!\frac{1}{2}
\left(1\!+\! \frac{1}{\sqrt{2^{n-1}}}\right)$ for $n$ odd, and
$P^{max}_C\!=\!\frac{1}{2} \left(1\!+\!
\frac{1}{\sqrt{2^{n-2}}}\right)$ for $n$ even. In the quantum
case, with the use of $n$ qubits in the maximally entangled (GHZ)
state, the task can be done with certainty for both cases, i.e.
$P^{max}_Q\!=\!1$. Note that in both cases in the limit $n\!
\rightarrow\! \infty$ the probability of success $P^{max}_C\!
\rightarrow \!\frac{1}{2}$ as by a simple random choice,
which drastically contrasts the certainty in the quantum protocol.

Next, suppose that the number $n$ of parties is {\it even} and
consider
$S'_{even}=\sqrt{2}\cos\left[\frac{\pi}{4}+(s_1+...+s_n)\frac{\pi}{4}\right]$.
This implies
\begin{equation}
g'_{even}=\sqrt{2^{n+1}} \cos\left[\frac{\pi}{2}(x_1\!+\!
x_2\!+\!...\!+\!x_n)+\frac{\pi}{4}\right].
\end{equation}
The success rate in the quantum protocol is now based on
violation of an inequality equivalent to the one of Ardehali
\cite{ardehali,belinskii}. The maximal probability of success in
a classical protocol is $P^{max}_C\!=\!\frac{1}{2}\left(1\!+\!
\frac{1}{\sqrt{2^{n}}}\right)$, whereas in the quantum protocol
with the use of the maximally entangled state the probability
reads $P^{max}_Q=\frac{1}{2}\left( 1+ \frac{1}{\sqrt{2}}\right)$.
Thus in this case one does not have certainty. However, because
the Bell inequality defined by $g'_{even}$ is violated by the GHZ
states by a higher factor than the one defined by $g_{even}$ (by
$\sqrt{2^{n-1}}$, instead of $\sqrt{2^{n-2}}$) the quantum
protocol is more resistant to the possible admixture of noise to
the GHZ states.

The advantage of the quantum protocol
associated with a given Bell inequality is a new measure of the
{\it strength} of such an inequality. The last example shows that this
measure favors GHZ-type contradictions for perfect correlations
($g_{even}$), making the factor by which the inequality is
violated less important (compare $g'_{even}$).

Let us look at generalizations. Consider $g$ as given by $\cos{(x_1+...+x_n)}$
where inputs $x_i$ belong to a continuous set $[0,2\pi)$ (while
$y_i$'s remain bits). The quantum protocol should now be adopted such
that each party has a choice to measure her qubit in a continuous
range of the settings of the apparatus. The protocol is based
on violation of the functional Bell inequality for continuous
range of the settings of the local apparatuses (for the
derivation and magnitude of violation see \cite{marek}):
\begin{equation}
\int_{x_1,...,x_n=0}^{2\pi} \cos{(x_1+...+x_n)} E(x_1,...,x_n)
\leq 4^n. \label{mzineq}
\end{equation}
One can apply the general proof obtained above if one replaces
$\sum_{x_1,...,x_n=0}^{1}$ by $\int_{x_1,...,x_n=0}^{2\pi}
dx_1...dx_n$ in the previous expressions. Thus, if and only if
the state violates the functional Bell inequality (\ref{mzineq})
the quantum protocol will have a higher success rate than the
classical one. The maximal probability of success in the
classical protocol is
$P^{max}_C=\frac{1}{2}\left(1+(\frac{2}{\pi})^{n-1}\right)$,
whereas in the quantum protocol with the use of $n$ qubits in
maximally entangled state this probability is
$P^{max}_Q=\frac{1}{2}\left(1+\frac{\pi}{4}\right)$. Similar
results can be obtained for an arbitrary  discrete number of
settings at each side.

For a next generalization, suppose that the inputs $y_i$ have $d$
possible values and that the two-valued function
$S[g(x_1,...,x_n)]$ is replaced with a function which has $d$
possible values. Then the quantum protocol should be based on the
violation of Bell's inequalities for $d$-dimensional quantum
systems (see \cite{collins}). Recently such a two-party protocol
was proposed \cite{brukner}.

We end with a remark. There are non-separable quantum states which
do not violate any Bell's inequality directly \cite{werner}. This
is often interpreted as implying that the violation of Bell's
inequalities is just a, even not optimal, entanglement witness,
without any significant importance for the implementation in
quantum information tasks. 
One cannot agree with such an interpretation
(see also Ref. \cite{scarani,acin}). The states which
only after local operations and classical communication  (LOCC) violate Bell's inequalities
\cite{popescu}, cannot be in any way useful in the communication complexity problems considered here.
Simply, any LOCC transformation requires {\it more} communication than it is  permitted
by the problems \cite{marek2}. 

M.\.{Z}. acknowledges KBN grant No. 5 P03B 088 20. The work is
supported by the Austrian  FWF project F1506, and by the European Commission, Contract-No. ERBFMRXCT96-0087,
and is a part of the Austrian-Polish program "Quantum Communication and Quantum Information IV" (2002-2003).


\begin{thebibliography}{99}
\bibitem{remark1} Note that, except
for shared entanglement, the discussed classical and quantum
protocols are performed under the same conditions. Obviously one
could consider wider classes of protocols which would include local calculations
of functions which depend on the full local inputs (i.e. on both $x_i$ and $y_i$).
However note, that the first term $y_1
\cdot...\cdot y_n$ in the function $f$ is completely random even
in the case when only one of the values $y_i$ is random, whereas
for the function $S[g]$ with inputs $x_1,...,x_n$ it is in
general not the case. This suggests that protocols with local computation of
functions which depends additionally on inputs $y_i$ might not be
of any advantage over the the discussed ones.
\end{thebibliography}
\end{document}